# Towards a 384-channel magnetoencephalography system based on optically pumped magnetometers

Holly Schofield[1,2], Ryan M. Hill[1,2], Lukas Rier[1,2], Ewan Kennett[1], Gonzalo Reina Rivero[1,2], Joseph Gibson[1], Ashley Tyler[2], Zoe Tanner[1,2], Frank Worcester[2], Tyler Hayward[2], James Osborne[3], Cody Doyle[3], Vishal Shah[3], Elena Boto[1,2], Niall Holmes[1,2] and Matthew J. Brookes[1,2]*

[1]Sir Peter Mansfield Imaging Centre, School of Physics and Astronomy, University of Nottingham, University Park, Nottingham, NG7 2RD, UK.

[2]Cerca Magnetics Limited, Units 7-8 Castlebridge Office Village, Kirtley Drive, Nottingham, NG7 1LD, Nottingham, UK

[3]QuSpin Inc. 331 South 104th Street, Suite 130, Louisville, Colorado, 80027, USA.

* Indicates correspondence to:
    Prof. Matthew Brookes,
    Sir Peter Mansfield Imaging Centre,
    School of Physics and Astronomy,
    University of Nottingham,
    Nottingham
    NG7 2UH
    United Kingdom
    E-mail: matthew.brookes@nottingham.ac.uk

Pages: 31

Words: 12,993

Figures: 9

Running title: Towards 384-channel OPM-MEG

**ABSTRACT**

Magnetoencephalography using optically pumped magnetometers (OPM-MEG) is gaining significant traction as a neuroimaging tool, with the potential for improved performance and practicality compared to conventional instrumentation. However, OPM-MEG has so far lagged conventional-MEG in terms of the number of independent measurements of magnetic field that can be made across the scalp (i.e. the number of channels). This is important since increasing channel count offers improvements in sensitivity, spatial resolution and coverage. Unfortunately, increasing channel count also poses significant technical and practical challenges. Here, we describe a new OPM-MEG array which exploits triaxial sensors and integrated miniaturised electronic control units to measure MEG data from up to 384 channels. We also introduce a high-speed calibration method to ensure the that the fields measured by this array are high fidelity. The system was validated using a phantom, with results showing that dipole localisation accuracy is better than 1 mm, and correlation between the measured magnetic fields and a dipole model is >0.998. We then demonstrate utility in human MEG acquisition: via measurement of visual gamma oscillations we demonstrate the improvements in sensitivity that are afforded by high channel density, and via a movie-watching paradigm we quantify improvements in spatial resolution. In sum, we show the first OPM-MEG system with a channel count larger than that of typical conventional MEG devices. This represents a significant step on the path towards OPMs becoming the sensor of choice for MEG measurement.

**INTRODUCTION:**

Optically pumped magnetometers (OPMs) are a newly emerging class of magnetic field sensor in the field of biomagnetism (see (Brookes et al., 2022; Schofield et al., 2023) for reviews). OPMs are small, lightweight, easy to use, and can measure magnetic fields with high sensitivity. Consequently, they are ideal for measuring the small magnetic fields generated by the human body. For magnetoencephalography (MEG) (measuring magnetic fields generated by the brain), OPMs have significant advantages over the more widely used superconducting quantum interference devices (SQUIDs). Specifically, 1) non-cryogenic sensors get closer to the scalp, measuring fields that are larger in amplitude and less spatially diffuse (Boto et al., 2016; Iivanainen et al., 2017)). 2) Lightweight sensors can be mounted in a helmet which moves with the head, enabling participant movement during a scan (Boto et al., 2018). 3) Unlike cryogenic sensors which are fixed in a "one-size-fits-all" helmet, OPMs can be mounted in helmets that accommodate multiple head sizes (babies to adults) (Hill et al., 2019). These advantages have been exploited in the study of babies (Corvilain et al., 2024) and neurodevelopment (Rhodes et al., 2025; Rier et al., 2024); to explore brain function underlying natural movement (O'Neill et al., 2025; Spedden et al., 2025); and to improve sensitivity (e.g. to epileptic spikes (Feys et al., 2022)). Commercial OPM-MEG systems can now be used "out of the box" (Demko et al., 2024) and offer exciting potential for neuroscientific and clinical applications.

Despite progress, OPM-MEG remains a new technology and even the most advanced systems make fewer than 200 independent measurements of the MEG signal around the scalp (i.e., <200 MEG channels) (Alem et al., 2023; Bonnet et al., 2025; Schofield et al., 2024). In contrast, conventional instruments have closer to 300 channels. This is important for three reasons. First, as sensors move closer to the head, field patterns become more focal (Boto et al., 2016; Iivanainen et al., 2017; Tierney et al., 2022); if the density of sensors is insufficient, then the largest magnetic fields can be missed (i.e., the field pattern is spatially aliased). This is particularly important for children, where the distance from the brain to the scalp surface (and sensors) is smaller than in adults (Boto et al., 2022) and field patterns become even more focal. Second, in most MEG studies fields from all channels are combined to create regional estimates of electrical activity in the brain – a process called source localisation. Increasing channel density affords an increase in the signal-to-noise ratio (SNR) of the reconstruction (Hill et al., 2024). Consequently, increasing channel count is key if OPM-MEG is to yield SNR values that are better than those achieved in conventional MEG. Finally, one significant advantage of OPMs is that their flexible placement allows coverage of regions that are hard to reach with conventional MEG – for example cerebellum (Lin et al., 2019). However, extending coverage to these areas, whilst maintaining coverage of the whole cortex, requires higher sensor counts. It follows that if OPM-MEG is to realise its potential as a high-fidelity imaging method, we must develop systems with a larger number of channels, surpassing the ~300 as is standard for cryogenic MEG instrumentation.

Maintaining the synchronous operation of a large array of OPMs is challenging. From a technical point of view, each OPM is a complex system in which a vapour of alkali atoms is used to sense local magnetic field. To enable measurement, the quantum mechanical properties of the atoms must be altered, both via absorption of laser light (optical pumping (Happer, 1972a)) and controlled manipulation of the magnetic fields experienced by the vapour. This requires operation of a semiconductor laser and a set of electromagnetic coils inside the sensor (to generate the required fields). The thermal properties of the vapour must be tightly regulated (Allred et al., 2002) as must the frequency of the light generated by the laser (which is also a function of temperature). Additionally, the magnetometer signal is read out via measurement of the amplitude of laser light passing through the vapour, requiring control of a photodiode (and lock-in detection). Successful OPM-MEG therefore requires these control signals to be sent, and output signals received, independently yet synchronously, to and from each of the sensors in the array. This necessitates complex electronics and the larger the array becomes, the larger the control system required.

Large arrays also pose significant practical problems. For example, source localisation requires that the locations of the OPMs and their sensitive orientations (relative to brain anatomy) are known precisely. This is straightforward with small numbers of sensors since optical methods can be used to track sensor locations. However, it becomes difficult with high-density arrays, not only due to increased numbers of sensors and their close proximity, but also because the sensors become obscured by cables. Source localisation also requires that the relationship between the output of a sensor (typically a voltage) and the magnetic field experienced by the sensor (the quantity required) is accurately known. This is termed the sensor gain and can vary from sensor to sensor (Hill et al., 2025); whilst it can be characterised by applying known fields to the sensor using on-board coils, running such an algorithm can be time consuming. In addition, when sensors are in close proximity, the field measurement made at one OPM can be affected by the presence of a second (an effect known as crosstalk); this problem becomes worse when using high-density arrays and can alter both directional sensitivity and gain. For these reasons, fast techniques to determine sensor locations, orientations, and gains – accounting for cross-talk – (henceforth termed array calibration) are critical for high-density arrays to be effective. Finally, once the array is calibrated, techniques to ensure that the array captures a faithful record of the real magnetic fields present (i.e., quality assurance techniques) are required. This is important for all studies and all array types, but becomes particularly critical if OPM-MEG systems are to be used clinically (e.g., for localisation of epileptogenic foci in pre-surgical mapping – a key application for high-density OPM-MEG systems (Feys et al., 2023; Rampp et al., 2020)).

In this paper, we aim to construct a 384-channel OPM-MEG system. We demonstrate how synchronised integrated miniaturised electronic control systems (Schofield et al., 2024) can be used to control a large-scale array of OPMs. Further, we develop a novel calibration methodology (Hill et al., 2025) which can accurately determine the positions, orientations and gains of the OPMs, and control for sensor cross talk. We employ a 'phantom' which generates well known "brain-like" magnetic fields to perform quality assurance checks on our array. Finally, we demonstrate the system with human MEG data, via the characterisation of task induced changes in neural oscillations during a visual task and the measurement of 'baseline' oscillations during a naturalistic 'movie watching' task. We hypothesise that higher channel count will show advantages for both sensitivity to neuromagnetic effects and spatial resolution.

**METHODS**

**OPM-MEG system design**

*Sensor array*

Our sensor array was constructed using 128 'triaxial' sensors (QuSpin Zero Field Magnetometers, Colorado, USA), with each sensor able to measure magnetic fields in three (notionally orthogonal) orientations with equivalent sensitivity (Boto et al., 2022). These three measurements are independent, and thus 128 triaxial OPMs provide 384 channels of data.

Each OPM contains a cell housing a $^{87}$Rb vapour. Circularly polarised 795-nm light (resonant with the D1 transition for $^{87}$Rb atoms) is passed through the cell and photons are absorbed by the atoms, changing both their energy levels and angular momentum. The result is to align the atomic magnetic moments along the direction of the laser beam and introduce a bulk magnetisation into the vapour (Happer, 1972b). Assuming no external magnetic field, the atoms become trapped in a single energy state; once in this state they can no-longer absorb photons and the amount of laser light passing through the vapour is maximised. However, in the presence of an external magnetic field, the bulk magnetisation interacts with that field according to the Bloch equations; this changes the energy states occupied by individual atoms and those atoms can once again absorb photons, causing a drop in light transmission through the cell. This system could act as a magnetometer, however the light passing through the cell (measured via a photodiode) is a Lorentzian function of magnetic field, peaking at zero. This makes it impossible to differentiate positive and negative fields. Thus, to add directional sensitivity, we use sinusoidally varying (923 Hz) modulation fields generated by on-board-sensor coils. This changes the solutions to the Bloch equations so sensor output (at the modulation frequency; measured via lock-in detection) is a linear function (distinct from a Lorentzian function) of the magnetic field through the cell, along the orientation of the modulation field (Cohen-Tannoudji et al., 1970) (i.e., if the laser beam is oriented in the z-direction and the modulation field is applied in x, we can measure the x-component of the external field). It is possible to measure two field components perpendicular to the laser beam (the x and y components) simultaneously by providing two temporally and spatially orthogonal modulation fields (oriented along x and y). However, the sensor is much less sensitive to fields along the beam axis (z-direction). Thus, to build a triaxial sensor, two beams are passed through the cell at a relative angle of 90°. The first beam, oriented in z, allows field measurements in the x and y directions. The second beam, oriented in x, allows field measurements in the y and z directions. The two beams combined then provide characterisation of all three components of magnetic field (Boto et al., 2022; Shah et al., 2020) (with the two measurements in the y direction combined). Importantly, to reach the high sensitivity required for MEG, the vapour must be heated to 150° enabling operation in the spin exchange relaxation free (SERF) regime (Allred et al., 2002) which maintains coherence between atomic magnetic moments.

We chose to build our array using triaxial sensors due to three significant advantages: first, having three-axes of field measurements (distinct from arrays with MEG sensitivity on a single axis (Alem et al., 2023; Hill et al., 2020); or two axes (Bonnet et al., 2025; Seedat et al., 2024) maximises the density of channels over a given region of scalp. This, in turn, maximises the total signal that can be measured by the array and consequently the SNR for source reconstruction (Hill et al., 2024). Second, triaxial measurement has been shown to improve the differentiation of magnetic fields from the brain and fields generated in the environment (Brookes et al., 2021; Holmes et al., 2023a). This is advantageous for interference rejection. Finally, triaxial measurement helps to improve the uniformity of coverage of the cortex (Boto et al., 2022), particularly for shallow areas, and especially in babies and children. Each triaxial sensor was 12.4×16.6×24.4 mm$^3$ and had an expected noise floor of 10-20 fT/√Hz along all three axes.

The sensors were mounted on the scalp via a 3D-printed bespoke helmet (Cerca Magnetics Limited, Nottingham, UK) which was designed based on the subject's MRI scan (Figure 1C). The internal surface of the helmet was made to fit the subject's scalp and face, whilst the external surface contained 163 approximately evenly spaced slots, into which the sensors could be located. The average distance from the scalp surface to the sensitive region of each sensor (i.e., the centre of the vapour cell, calculated for all available helmet slots) was 6.2 ± 5.0 mm (mean ± standard deviation). Because the helmet was based on the subject's MRI, the sensor locations relative to brain anatomy were known a-priori from the computer aided design (CAD) files of the helmet and sensors.

*Electronics*

The sensors were controlled by two integrated miniaturised electronics units ((Schofield et al., 2024) *Neuro-1-electronics*, QuSpin, Colorado, USA – https://quspin.com/neuro-1-an-integrated-sensor-system-for-opm-meg/) working in concert; with each unit controlling 64 OPMs.

Within a single unit, each OPM is controlled by a separate circuit board, which provides control signals for all on-board sensor components and enables lock-in detection to read the output signals. Four signals are initially read from the two photodiodes in the sensor (these reflect fields in the x, y (measured twice), and z orientations). The card then outputs three *digital* signals (which are directly proportional to magnetic fields along the three sensor axes) to a multiplexer. Cards are grouped together in modules, with each module containing 8 circuit boards to control 8 sensors. Each electronics unit houses 8 modules to control a total of 64 sensors. The outputs from each module are combined using the multiplexer and are then sent to a network card. Each electronics unit is 0.36 x 0.2 x 0.06 m$^3$ and weighs 0.81 kg. Importantly, to make the two units work together, they require a synchronous clock. This was provided by a signal generator which made a 12.288 MHz, 1.8 V (peak-to-peak, positive voltage only) sine wave, which was fed to both units. In theory, it would be possible to combine any number of units/modules/cards in this way so the total channel count could be increased (or decreased) in accordance with requirements for any MEG experiment.

Both electronics units then connected via Ethernet to a custom data acquisition system (DAQ) based on a FPGA controller — sbRIO9637, National Instruments, which integrates the MEG signals from all 384 channels and combines them with external inputs representing stimulus timing. All signals are then synchronously sampled and passed to an acquisition PC via an ethernet connection for visualisation and storage. A schematic illustration of the full system is shown in Figure 1A, along with a schematic of the individual IM control units in Figure 1B.

As an aside, the two units are also capable of three-axis closed-loop operation (in which the magnetic field is measured at each OPM along all three axes, and the electronics effects a negative feedback loop, whereby currents are applied to the on-board sensor coils to maintain zero field at the vapour cell). This has been shown to linearise the OPM response in cases where background field is high (Schofield et al., 2024). However, it wasn't employed for the current work as large dynamic range wasn't required. In addition, the decoder also has 16 analogue to digital converter channels and 9 digital inputs. These auxiliary channels can be used to sample additional signals synchronously with OPM data (e.g., signals from stimulus equipment).

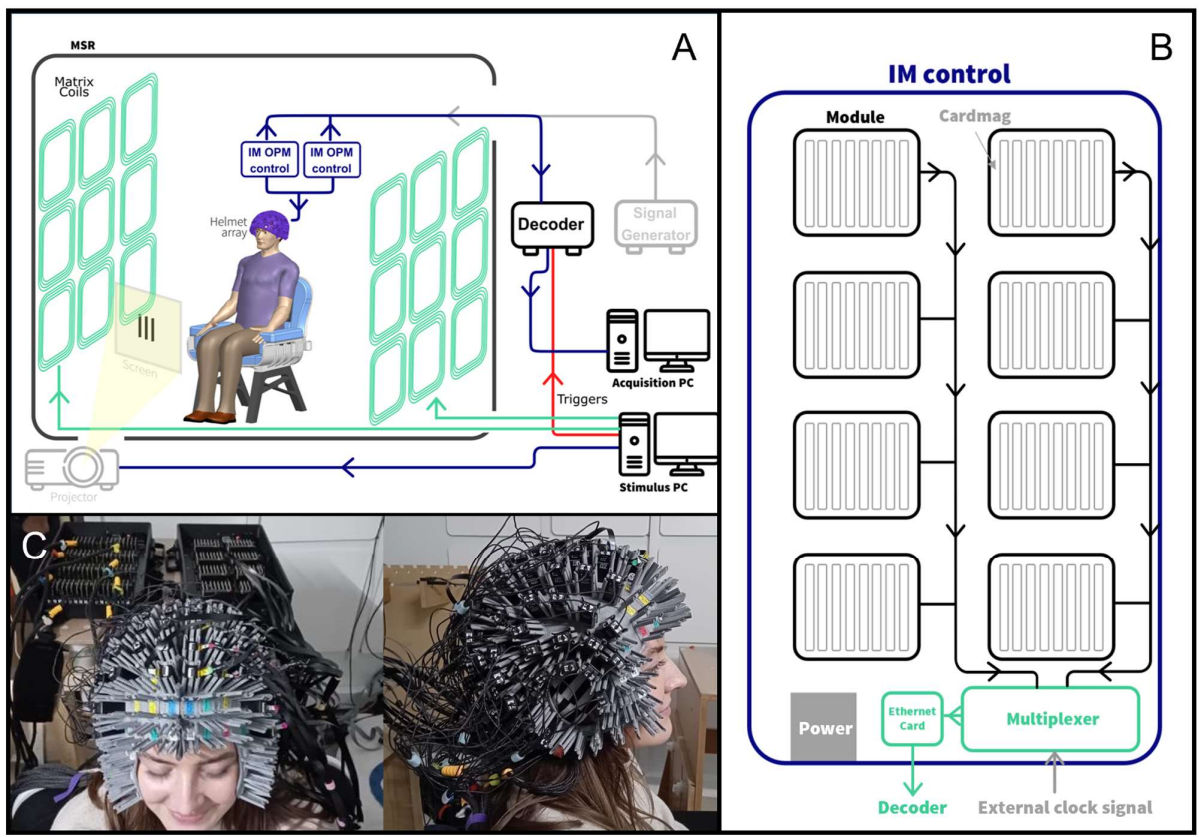

*Figure 1: The OPM-MEG system:* A) Schematic diagram of the complete OPM-MEG system. B) Schematic of an individual IM control electronics. C) 384 channel triaxial OPM-MEG array, with sensors mounted close to the scalp using a bespoke 3D printed helmet.

*Magnetic shielding*

The sensor array and electronics units were housed inside a magnetically shielded room (MSR) comprising four MuMetal layers to attenuate DC/low frequency and one copper layer to attenuate high frequency magnetic interference fields (MuRoom, Magnetic Shields Limited, Kent, UK). The MSR walls were equipped with degaussing coils to reduce remnant magnetisation prior to a scan. The MSR was also equipped with a matrix coil (Holmes et al., 2023b) capable of active field control (Cerca Magnetics Limited, Nottingham, UK). Active field compensation was not used for these experiments. However, the matrix coils were used to provide fields for external calibration (see below).

*System control*

A single "acquisition" computer was used for OPM-MEG control, degaussing, and data acquisition; the experimental paradigms (along with associated temporal markers delineating the time at which stimuli were presented to the subject – termed triggers), phantom control and matrix coil calibration procedures (see below) were controlled by a second "stimulus" computer. The MSR was equipped with a projector to present visual stimuli into the MSR, via projection through a waveguide onto a back projection screen positioned ~100 cm in front of the subject. We used an Optoma HD39 Darbee projector (refresh rate 120 Hz).

**System calibration**

System calibration involves finding of the locations of the sensors (relative to each other), the orientations of their sensitive axes and the gain of each of each channel. We did this by imposing a set of known magnetic fields onto the array, and fitting a model of those fields to the measurements made at the sensors (Hill et al., 2025; Iivanainen et al., 2022).

To generate the fields, we used the matrix coil which comprised 96 independently controllable coil elements (Holmes et al., 2023a) spaced approximately evenly on all 6 faces of the MSR. The field per unit current generated by each coil element had previously been mapped throughout the centre of the MSR with high precision, using a fluxgate magnetometer (Hill et al., 2025). Using this mapping, we determined the optimum combination of coil currents to create three uniform fields oriented in $x$ ($B_x$), $y$ ($B_y$), and $z$ ($B_z$). In addition, we determined the optimum combination of coil elements to generate the 5 independent magnetic field gradients; $\frac{dB_x}{dx}, \frac{dB_x}{dy}, \frac{dB_x}{dz}, \frac{dB_z}{dz}, \frac{dB_y}{dz}$.

The coils were operated so each of the 8 fields were generated, simultaneously, by currents fed to the matrix coils at different frequencies (3-10 Hz in steps of 1 Hz) (i.e. a field oscillating a 3 Hz represented $B_x$, a field at 4 Hz represented $B_y$ and so on). We generated all 8 fields simultaneously for a period of 4 s. We then took a segment of the data and, for every channel, used a Fourier transform to quantify the direction and magnitude of the measured field at each of the 8 frequencies.

We used the three uniform fields to determine the orientation and gain of each channel. Specifically, we modelled the signal measured by a single channel as

$$\boldsymbol{b} = \boldsymbol{B_u}\boldsymbol{g}. \qquad [1]$$

Here, $\boldsymbol{b}$ represents a 3-element column vector corresponding to the magnitude of the 3 uniform field components, measured by a single channel. $\boldsymbol{g}$ represents the (unknown) sensor gain along three orthogonal axes and $\boldsymbol{B_u}$ is a diagonal square matrix whose on diagonal elements represent the known magnitude of the three uniform field components ($B_x$, $B_y$ and $B_z$) generated by the matrix coil, which for our purposes were set to 0.1 nT along the $x$, $y$ and $z$ directions. We then solved the equation for $\boldsymbol{g}$ ($= \boldsymbol{B_u^+}\boldsymbol{b}$; where $\boldsymbol{B_u^+}$ is the pseudo-inverse of $\boldsymbol{B_u}$); normalisation of $\boldsymbol{g}$ gives sensor orientation, and its magnitude gives the channel gain.

Following this, we used the 5 field gradient terms, alongside knowledge of the channel orientation and gain, to solve for the channel location. Specifically, we can write,

$$\boldsymbol{b_g} = \boldsymbol{B_G}\boldsymbol{r}, \qquad [2]$$

where $\boldsymbol{b_g}$ is the 5-element column vector representing the signal measured at each channel due to each of the 5 gradients. $\boldsymbol{B_G}$ is a 5x3 matrix, where a single column characterises the vector components of field that would be measured, based on the estimated channel orientation and gain, if the channel were at a distance of 1 m from the origin of the coordinate system (the centre of the helmet) in all three cartesian directions. (I.e. column 1 represents the fields for a $\frac{dB_x}{dx}$ gradient; column 2 the $\frac{dB_x}{dy}$ gradient, and so on.) Multiplying this matrix by the 3-element column vector $\boldsymbol{r}$ (which represents the real sensor locations) generates a model of the measured gradient fields. To solve for $\boldsymbol{r}$ (i.e. to obtain the channel location) we again use the pseudo-inverse approach, computing $\boldsymbol{B_G^+}$. For each sensor (i.e. three orthogonal channels) we took the position to be the average of the three channel positions. By repeating this process for every channel (sequentially) we derive a complete description of the locations, orientations and gains of every channel in the array.

(Iivanainen et al., 2022) used this same approach (with 18 coil elements) to produce a starting guess for a more advanced spherical harmonic refinement (based on matching channel responses to spherical harmonic models of the fields produced by each coil). Here, the larger number of coil elements ostensibly enables more accurate generation of field variations, and removes the need for harmonic refinement. Our own previous work (Hill et al., 2025) used a 'coil-by-coil' approach (whereby signals from 96 coils were fitted independently). However, this required 20 s of data to spectrally resolve signals from all 96 elements. Here, by condensing to just 8 signals, the acquisition time can be dramatically reduced (see also Appendix).

**Phantom experiments**

To test the accuracy of our 384-channel array (and its calibration) we used an electromagnetic phantom (Cerca Magnetics Limited, Nottingham, UK) which generates controlled 'brain-like' magnetic fields. The phantom was constructed from a semi-circular printed circuit board (PCB) containing 5 magnetic dipoles. Each dipole comprised a spiral coil fabricated on a two layer PCB, with 20 turns on each layer connected by a via at the spiral's centre. The spiral's inner radius was 1.4 mm with an outer radius of 6.3 mm. The dipoles are positioned on a circular path, at a radius of 7.5 cm, at intervals of 30°. Each dipole is supplied with current via wire paths on the PCB connected to a twisted pair cable, which runs from the phantom to the outside of the MSR. This was connected to a voltage source (National Instruments NI-9264 16-bit digital-to-analogue converter) with a 56-kΩ resistor in series, and controlled using MATLAB.

The phantom was placed at the centre of the MSR inside a 3D printed helmet with the dipoles running approximately along the anterior-posterior direction. (Note, the helmet used to accommodate the phantom was slightly larger than that used for the human study, but similar in design.) Each dipole was energised sequentially: Within a single "trial", the dipole was energised for 2 s using an oscillating driving voltage at 27 Hz, with a peak-to-peak amplitude of 1 V. This was followed by a 1-s window with the dipole switched off. We repeated 50 trials of the experiment for each dipole (total experimental time 150 s). Data were recorded from all OPM channels throughout. This process was repeated for all 5 dipoles. We repeated the whole experiment twice.

*Phantom data analysis*

Data were segmented and averaged across trials, and a single timepoint was selected at one of the peaks of the trial averaged dipole timecourse. Extracting data from this single time point for all channels, and repeating this process for each dipole, resulted in 5 magnetic field patterns representative of each of the 5 dipoles. We then used a dipole fitting algorithm to determine the location, orientation, and amplitude of each dipole. Models were created based on the equation for the field produced by a magnetic dipole, and fitted using MATLAB's non-linear regression function *nlinfit* (minimising the difference between the measured and modelled fields). We then undertook two analyses to determine the accuracy of the result:

1) **Dipole localisation errors**: Having determined the locations of the 5 dipoles, we computed the distances between all possible dipole pairs (there are 10 measurable distances between locations). We then compared these distances to their known values from the PCB manufacturing process (using the centre of the spirals). This enabled quantification of the localisation error.

2) **Correlation between measured and modelled fields**: For each of the 5 dipoles, we measured the Pearson correlation coefficient between the measured and modelled magnetic field patterns.

The above analyses were repeated twice; once where we used the sensor locations, orientations, and gains derived using calibration, and a second time where sensor locations and orientations were taken directly from the computer aided design (CAD) file for the helmet – this assumes that channel orientations are orthogonal and parallel to the outer casing of the sensor, and that all sensor gains were unity. We hypothesised that distance errors would be lower and correlations higher when using calibration.

**Human experiments**

*Experimental Paradigms and data recording*

To test the suitability of our system for recording human MEG data, a single participant undertook a series of experiments. The participant provided written informed consent prior to taking part, and all experiments were approved by the University of Nottingham School of Medicine and Health Sciences

Research Ethics Committee. The participant was sat comfortably on a support at the centre of the MSR, wearing the bespoke helmet (Figure 1C). We used two different experimental paradigms:

- **Visual experiment:** Visual stimulation was applied using a circular grating, displayed centrally at a visual angle of 24°, with 3 cycles per degree. A single trial comprised 2 s of stimulation where the circular grating was static, followed by 2 s where the grating was allowed to drift inwards at a rate of $2°s^{-1}$ grating. This was followed by a rest period of 6 s during which a white fixation cross was located centrally on a black screen. Sixty trials were shown, making the total experimental duration 600 s. Stimulus timing was sent from the stimulus PC to the OPM-MEG decoder via a parallel port. On cessation of visual stimulation, the subject was asked to make a finger abduction with the index finger of their right hand. This stimulus generates both a task induced reduction in alpha oscillations and an increase in gamma oscillations, in primary visual cortex (Fries et al., 2008).
- **Movie watching experiment:** The participant watched a clip from the film 'Dog Day Afternoon' (as also used in (Rier et al., 2023)). Data were recorded for 622 ± 8 s (mean ± standard deviation across runs). Here, we aimed to characterise baseline oscillatory power across brain regions.

During both experiments, the participant was free to move but not told to do so. In a single recording session, the subject undertook both experiments. The session was repeated five times. In all cases the OPM-MEG data were recorded at a sample rate of 375 Hz. Following collection, all data were organised according to the Brain Imaging Data Structure (BIDS) for MEG (Niso et al., 2018). The participant also underwent an anatomical MRI scan (1-mm isotropic spatial resolution; T1 contrast; 3 T Phillips Ingenia MRI scanner).

*Visual task: Data preprocessing*

All data processing was carried out using the Fieldtrip toolbox (Oostenveld et al., 2011) alongside custom written code in MATLAB. A notch filter (50 Hz) was applied to all OPM-MEG data to remove powerline noise and its harmonics. Data were then bandpass filtered into the 1 – 150 Hz band using a 4th order Butterworth filter. A power spectral density plot was created (using Welch's method (Welch, 1967)) for all channels, and any channels with no signal (<7 fT/√(Hz)) or showing high noise (>40 fT/√(Hz)) in the 60 – 80 Hz range were removed (based on visual inspection). Homogeneous field correction (HFC) (Tierney et al., 2021) was applied to all data to remove interference that manifests as a spatially uniform field across the OPM array. OPM-MEG data recorded were segmented into 10-s trials (0 s < t < 10 s relative to visual stimulus onset). Any bad trials (defined as an individual trial in which the signal variance (at any one channel) exceeds three times the standard deviation of variance measured across all trials (for that channel)) were removed. All trials were also inspected visually and any remaining trials with high noise levels were removed.

*Visual task: gamma band oscillations*

We used a beamformer (Robinson & Vrba, 1998) to construct pseudo-T-statistical images showing the spatial signature of task induced gamma band power change. Data were filtered to the 40-60 Hz band and a data covariance matrix constructed using data from the whole experiment (excluding bad trials). Tikhonov regularisation was applied to the matrix (using a regularisation parameter equal to 1% of the maximum eigenvalue of the unregularized matrix). The brain volume was divided into voxels on a 4-mm cubic grid and the beamformer weighting parameters (which define the linear combination of channels that optimally describe the electrophysiological signal at each voxel) were constructed, using the regularised data covariance matrix and a forward solution based on a single shell volume conductor model (Nolte 2003). To make the pseudo-T-statistical image, we contrasted projected oscillatory power in the 2 s to 4 s (active) window to oscillatory power in the 7 s to 9 s (control) window (timings relative to onset of visual stimulation). Pseudo-T-statistics were derived for all voxels, with source orientation determined as the direction of maximum beamformer projected signal amplitude (Sekihara et al., 2004). The resulting Pseudo-T-statistical images were then averaged across the 5 experimental runs.

For the location showing the largest gamma response (derived individually for each run), we generated a time-frequency spectrogram (TFS). Beamformer weights were derived using a covariance matrix calculated in the 1 – 150 Hz band and regularised as above. This, and the forward model, were used to derive a broadband estimate of the timecourse of electrophysiological activity at the location of interest (i.e. a virtual electrode (VE)). VE data were then frequency-filtered into overlapping bands between 1 Hz and 120 Hz; for each band, a Hilbert transform was used to derive the analytic signal, and the absolute value of the analytic signal was calculated to generate the instantaneous amplitude of the band-limited oscillations – termed the Hilbert envelope. This was averaged over trials and concatenated across frequencies. For all frequency bands, we calculated a baseline oscillatory amplitude (in the 8 s ≤ t ≤ 9 s time window). We then computed the final TFS as the difference between instantaneous oscillatory amplitude and baseline, normalised by the same baseline to give relative change in oscillatory amplitude throughout the task trial.

*Visual task: the effect of high channel count*

In addition to the TFS, we derived the trial averaged gamma band Hilbert envelope in the 40 Hz to 60 Hz frequency range. We then computed SNR as the mean difference in Hilbert envelope between the active and rest windows, divided by standard deviation of the envelope in the rest window. To investigate how channel density affects SNR, we recomputed this same gamma envelope with a random selection of channels in the array turned off. The number of channels used ranged from 1 to 331, in steps of 10, with 10 iterations of random channel distributions for each channel count. To assess the effect of varying the number of channels, we plotted SNR as a function of the total measurable signal, which was quantified as the Frobenius norm of the forward field vector, $\|l\| = \sqrt{\sum_{i=1}^{N} l_i^2}$ (where $l_i$ is the forward field for channel $i$ and $N$ is the total number of channels used). Note that $\|l\|$ changes with the square root of the total channel count (Hill et al., 2024). For these calculations, we computed a gamma envelope with no HFC applied to allow the full range of channel counts to be used. (At the lower end of the range, there would not be sufficient channels to perform HFC).

*Movie watching task: Preprocessing*

Data processing for the movie watching task was carried out using MNE-python (Gramfort et al., 2013). Notch filters were applied at the powerline frequency (50 Hz) and 2 harmonics, and data were filtered into the 5-150 Hz band. As previously, PSD plots were created, and channels with low or excessively noisy signals were removed. An automated outlier detection procedure adapted from OSL-ephys (van Es et al., 2025) was used to identify and remove data segments with high noise. All data were also inspected visually, and any remaining segments with high noise removed. HFC was applied to remove interference.

*Movie watching task: Spectral Power*

We parcellated the brain into distinct anatomical regions according to the Automated Anatomical Labelling (AAL) atlas. This was achieved by co-registering the AAL atlas to the participants anatomical MRI scan, using FLIRT in FSL (Jenkinson et al., 2002; Jenkinson & Smith, 2001). We found the location of the medoid voxel of each AAL region and generated a VE timecourse at those locations. To calculate the VEs, beamformer weights were derived using data covariance computed in the 5 – 150 Hz band, and a time window encompassing the entire experiment (excluding bad data segments). The forward model was a single-shell volume conductor (Nolte, 2003). The covariance matrix was regularised, using a regularisation parameter equal to 5% of the maximum eigenvalue of the unregularized matrix. This process resulted in 78 timecourses, representing each of the AAL regions.

For each regional timecourse, we used Welch's method to estimate the broadband PSD. In addition, we frequency filtered the VE data to both the alpha (8 – 13 Hz) and beta (13 - 30 Hz) bands (both using a $4^{th}$-order Butterworth filter) and derived two more PSDs. We then computed the area under the alpha/beta band PSDs and normalised this by the area under the broadband PSD, to give a measure of "relative alpha/beta power" (i.e., the fraction of the total spectrum that exists in the alpha or beta band). This was repeated for all brain regions and plotted as a functional map. We hypothesised that, similar to previous literature (Hunt et al., 2016), relative alpha power would peak in the occipital regions whilst relative beta power would be a maximum across the parietal lobes and sensorimotor regions.

*Movie watching task: Spatial Resolution*

Finally, we aimed to assess the effect of high channel count on our movie watching data, reasoning that increasing the number of channels would improve spatial resolution. In this context, lower spatial resolution means higher leakage of signal between the regional timecourses. To quantify this, for every possible pair of brain regions in the AAL parcellation, we quantified zero-phase-lag signal leakage between locations using linear regression (Brookes et al., 2012) (this effectively measures the shared variance between brain locations, which becomes larger for lower spatial resolution). This was plotted as a matrix, where the element $[i,j]$ represents the source leakage between each pair of regions, $i$ and $j$. We then averaged these matrices across columns, yielding a 78-element vector whose elements represent the total leakage from a single region to all other regions. Having computed this for the data reconstructed using the full (384-channel) array, we then randomly selected 192 channels and recomputed the VE timecourses and the leakage. This process was repeated 20 times to obtain multiple realisations of the (randomised) 192-channel array. We hypothesised that leakage would be lower for the 384-channel array compared to the 192-channel arrays.

**RESULTS**

**System performance, calibration and phantom data**

Figure 2A shows the noise level of the sensors when operated as a high-channel-count system. Power spectral density is overlaid for all channels, and the dashed black line represents a 15 fT/√Hz level. We removed 38 channels, either due to them measuring no signal, or having high noise levels. Of the remaining 346 sensors, the median noise in the 3 to 100 Hz frequency band was 17 fT/√Hz, which is consistent with existing literature (Boto et al., 2022; Schofield et al., 2024) for triaxial sensors. Figure 2B shows the channel layout derived using the calibration process. The direction of each arrow represents the channel orientation, and the colour represents the calibration derived gain value.

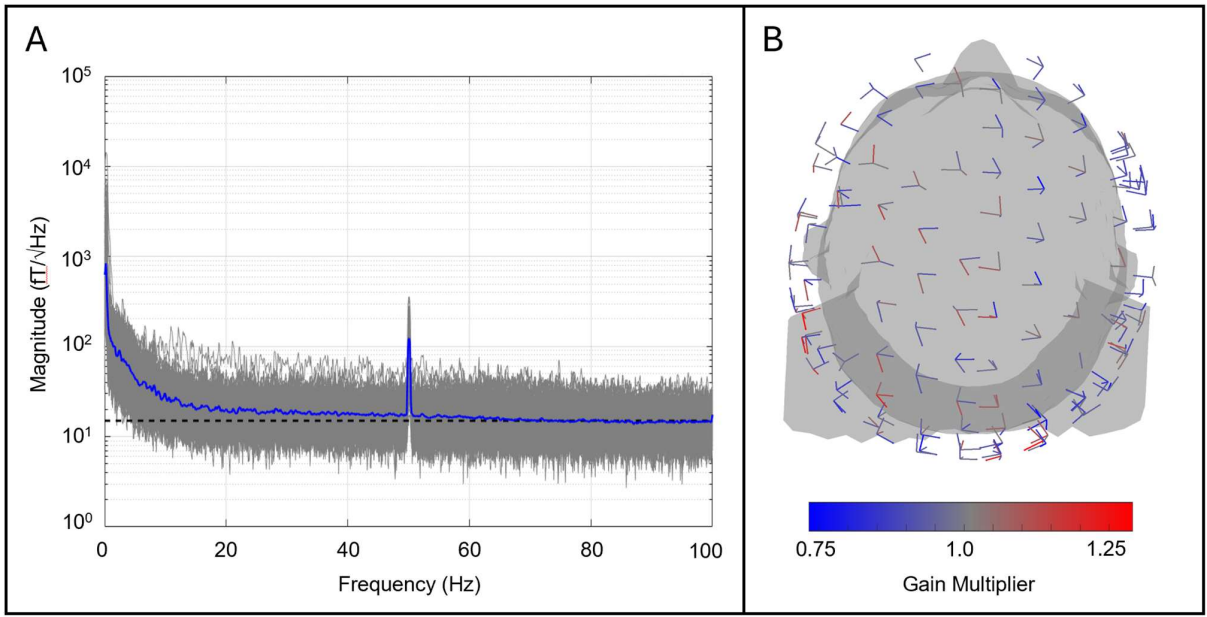

*Figure 2: Sensor performance and calibration. A) Shows the power spectral density of empty room noise, measured for 346 channels and overlaid. The dashed line is at 15 fT/sqrt(Hz), and the blue line shows the mean of all channels. B) Calibration derived channel locations, orientations and gains. Each channel is represented by an arrow which denotes the sensitive orientation. The point where multiple arrows converge is the location of the sensitive volume of a sensor. Channel gains are presented by the colour of the arrow.*

Figure 3 shows results of the phantom study. In Figure 3A, the grey area represents the size/shape of the phantom PCB, and the blue circles depict the known locations (from the PCB manufacture) of the 5 magnetic dipoles. The crosses show the dipole locations derived using dipole fitting applied to the recorded magnetic field data (only data from the first experimental run are shown). Recall we ran the analysis twice, with and without calibration; the blue crosses show the locations derived with calibration and the red crosses show the locations derived with no calibration. Note that qualitatively, there is good agreement between the dipole fitting process and the underlying PCB structure.

To quantify the error in dipole localisation we measured the distances between all possible pairs of fitted locations and then compared the results to the equivalent distances measured via the PCB manufacture (the latter representing a 'ground truth'). The matrices in Figure 3B show the resulting distance errors (e.g., element [1,2] represents the error in the distance between dipole 1 and dipole 2, recorded during the dipole fitting process). Results were derived with (left) and without (right) calibration; data are shown for the first experimental run only. These distance errors are further quantified in Figure 3C which shows the mean and standard deviation across all dipole pairs, with (blue) and without (red) calibration. As shown, with calibration localisation errors are less than 1 mm for both experimental runs (0.89 ± 0.5 mm for run 1 and 0.87 ± 0.4 mm for run 2). Without calibration these errors increased (to 1.3 ± 1.1 mm for run 1 and 2.0 ± 1.4 mm for run 2). Figure 3D shows the correlations between the magnetic field patterns measured by the OPM array, and the fields predicted by a magnetic dipole model. With calibration, the average correlation is 0.9985 ± 0.0002 for run 1 and 0.9984 ± 0.0009 for run 2. Without calibration, this correlation drops to 0.98 ± 0.02 for run 1 and 0.98 ± 0.01 for run 2. Both results highlight the advantages of calibration.

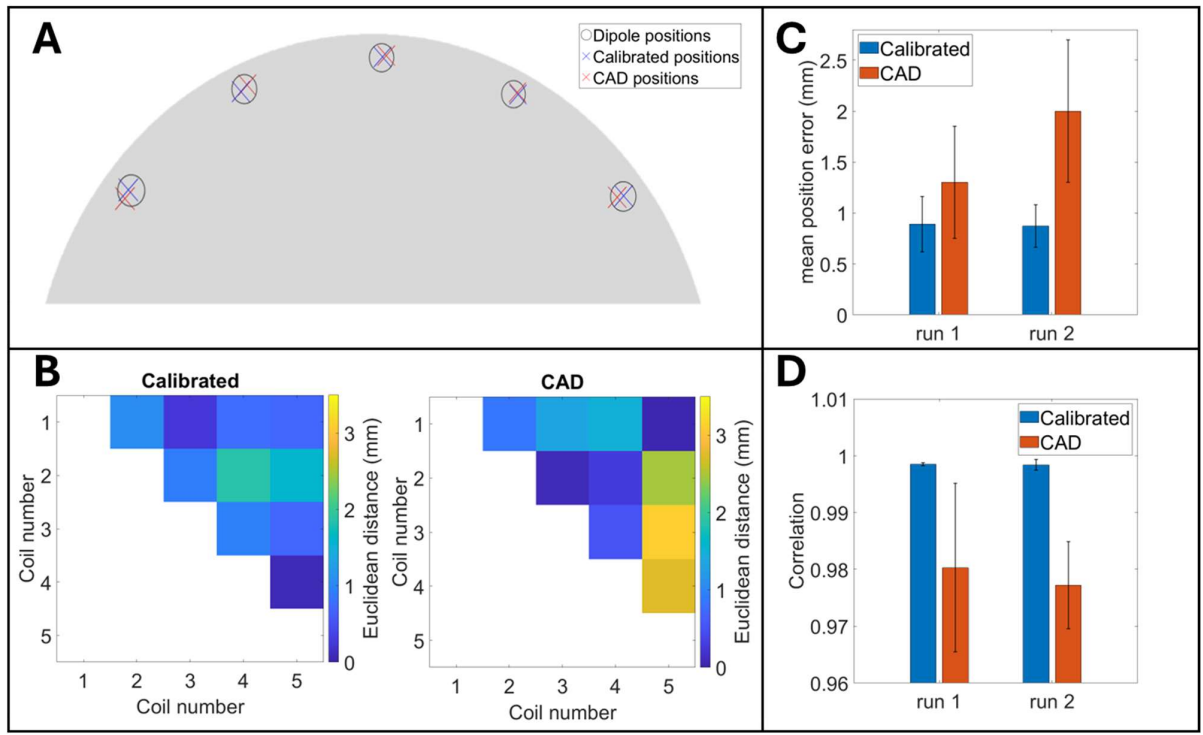

*Figure 3: Phantom results.* A) a graphic representation of the phantom (grey semi-circle) with the ground truth dipole locations shown by the blue circles. The blue crosses show the dipole locations derived using a dipole fit to the measured magnetic field data, using calibrated sensors. The red crosses show positions derived from the same process, but using the CAD model for array geometry and assuming unit gain (i.e. no calibration). Data presented are from Run 1 only. B) Discrepancies in the distance between dipole pairs, measured using dipole fitting and directly from the PCB manufacture.

*Left hand plot shows with calibration and right-hand plot shows without calibration. Data from Run 1 only. C) Dipole localisation errors, quantified as the mean and standard deviations of the elements of the matrices in (B). Data from both experimental runs are shown. D) Corelation values between the measured and modelled magnetic fields from the phantom. Note that with calibration, dipole fitting errors are <1 mm and correlation values are >0.998.*

**Human experiments: Visual gamma results:**

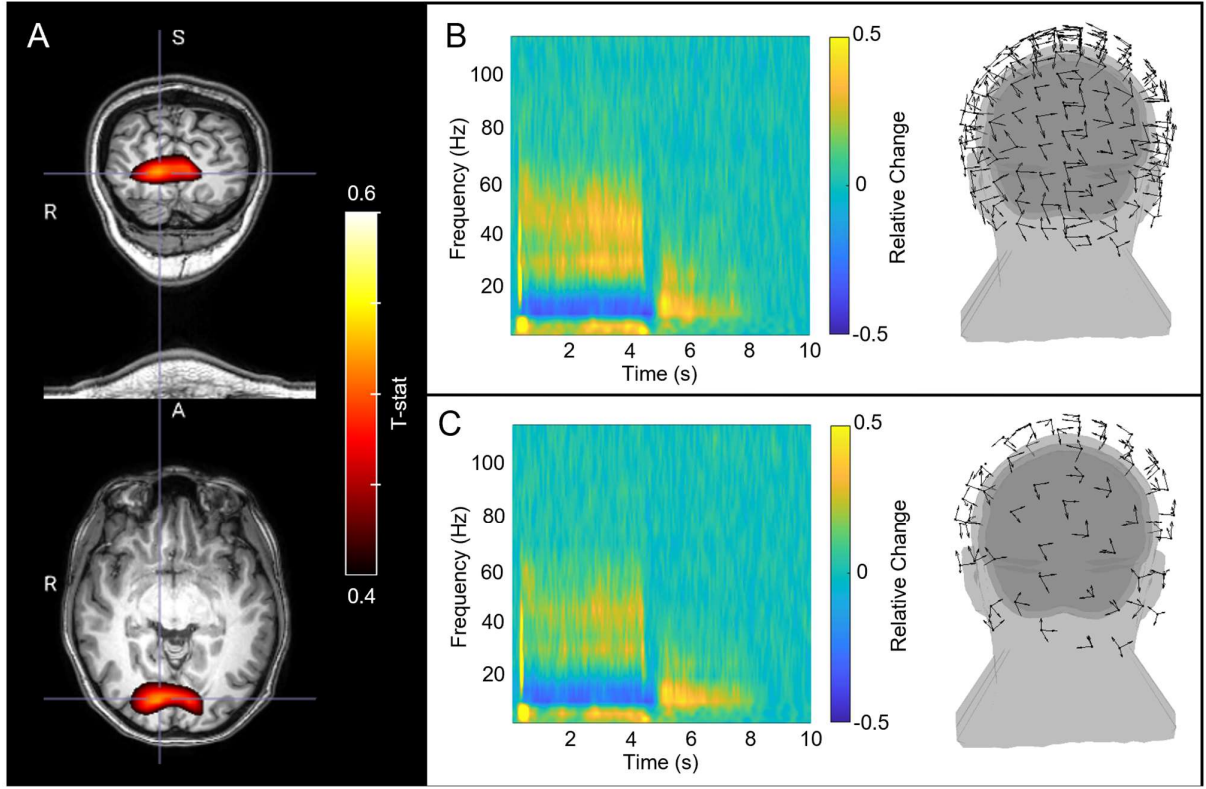

*Figure 4: visual gamma results: A) The spatial distribution of gamma power change with visual stimulation (red/yellow) overlaid onto the subject's anatomical MRI scan (computed using the full array). B) TFS data from the location of maximum stimulus-induced gamma power change; reconstruction from the full array (layout shown). C) TFS data from the same location reconstructed using a sparse array with 174±5 channels (again, array layout shown).*

Figure 4 shows results from the visual experiment. On average, across the 5 runs of the experiment, we had 349 ± 3 working channels. On average we removed 6.2 ± 0.7 trials due to artifacts in the data. In Figure 4A, the red/yellow overlay shows the spatial signature of stimulus induced change in gamma power, captured using the full array. This is averaged across experimental repeats and overlaid onto the subject's anatomical MRI scan. The location of the largest change is in primary visual cortex, extending to both hemispheres, which is consistent with a centrally presented visual stimulus. Figure 4B shows a TFS representation of the VE data extracted from the location of largest gamma response. Yellow represents an increase in oscillatory amplitude relative to baseline; blue represents a decrease. Data are reconstructed using all channels and the array layout is shown. The result is again averaged across all 5 experimental repeats. Figure 4C shows the same result, but the number of channels has been reduced to 174± 5 (a subset of 64 evenly distributed sensors after bad channel removal for each run) and the TFS re-derived (the layout of the sparse array is again shown). In both cases, a clear increase in narrow band gamma oscillations at approximately 45 Hz is apparent and this is consistent with literature using similar paradigms (Fries et al., 2008; Muthukumaraswamy, 2013). However, by reducing the channel count we see a drop in the SNR of the gamma signal. Quantitatively, by moving from 349 to 174 channels, the total possible measurable signal (quantified as the Frobenius norm of the forward field pattern from the location of maximum gamma change, $\|l\|$) reduced from 67 ± 4 fT to 46 ± 9 fT (mean and standard deviation across experimental runs). The experimentally derived SNR of the gamma response decreased from 5.8±1.0 to 4.6±0.8.

Figure 5A further quantifies the advantages of a high channel count by plotting SNR of the gamma response as a function of the total measurable signal. As noted in our Methods section, the total measurable signal is measured as the Frobenius norm of the forward field pattern from the location of interest in visual cortex, $\|l\|$. Here this quantity is divided by $\|l_{max}\|$, which represents the same Frobenius norm calculated assuming all 163 available sensor in the helmet slots had been filled. I.e, a value of $\frac{\|l\|}{\|l_{max}\|} = 0.8$ denotes that we are measuring 80% of the signal that we could expect from a fully populated helmet. Data from all 5 experimental repeats are shown. As expected, (Hill et al., 2024) there is an approximately linear relationship between $\frac{\|l\|}{\|l_{max}\|}$ and SNR, meaning SNR is improved with channel count.

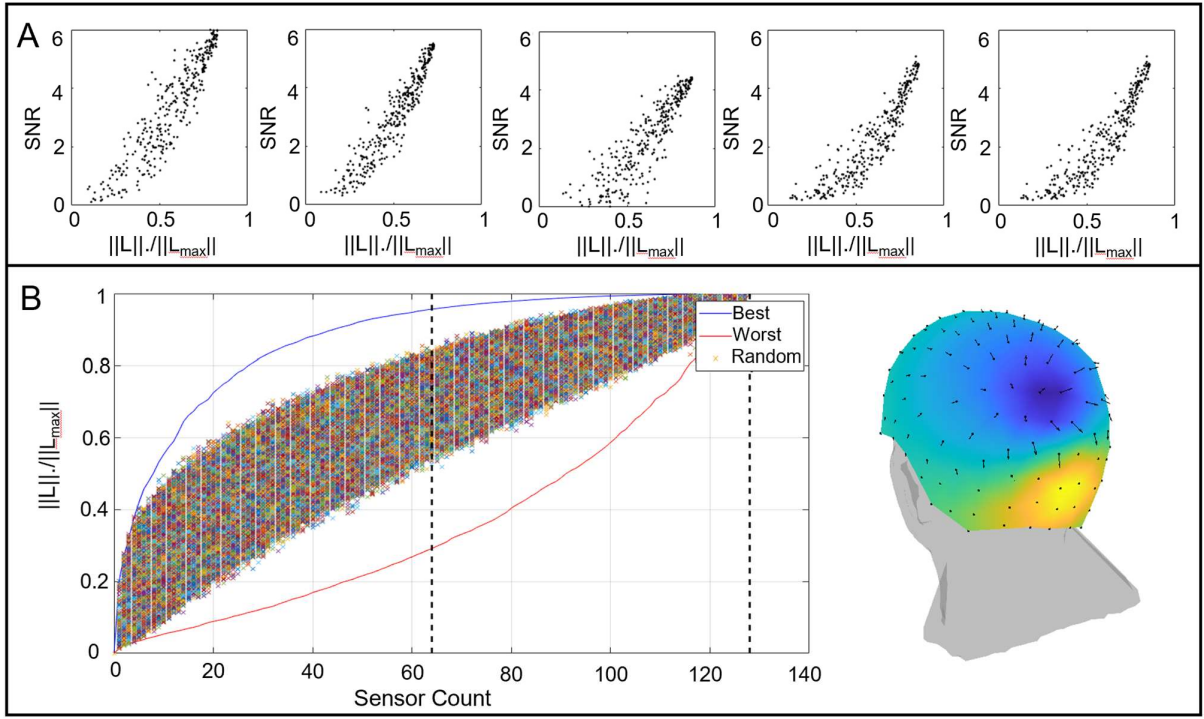

*Figure 5: Channel selections:* A) SNR as a function of the total measurable signal (which is quantified as $\frac{\|l\|}{\|l_{max}\|}$ and varied based on different channel counts). All 5 experimental repeats shown separately; note the linear relationship. B) Total measurable signal for a dipole in visual cortex (field pattern shown inset) as a function of channel count. The crosses show the case where different numbers of sensors are switched on/off randomly. The blue line shows the case where sensors switched on are located judiciously over visual cortex. The red line shows the case for poorly positioned sensors.

However, increasing channel count is not the only way to change $\|l\|$ and SNR. Figure 5B shows $\frac{\|l\|}{\|l_{max}\|}$ plotted as a function of the number of sensors used; the plot is shown for the specific case of a current dipole

in visual cortex (at the location of the largest gamma change – the field pattern generated by this dipole is also shown). The crosses show values of $\frac{\|l\|}{\|l_{max}\|}$ derived when sensors were selected at random and switched off; it's clear that $\frac{\|l\|}{\|l_{max}\|}$ increases (approximately) with the square root of channel count, as would be expected. However, the blue line shows the case where sensors are judiciously located to maximise the measurable signal (i.e. sensors are strategically paced over visual cortex). Here, $\frac{\|l\|}{\|l_{max}\|}$ (and SNR) increases much faster, and we can get within 5% of the highest possible signal with just 59 sensors. For completeness the red line shows the case where sensors are poorly located. This shows that to achieve high sensitivity using an OPM-MEG array we don't necessarily require a high channel count, if we can optimally distribute sensors over brain regions of interest. This will be addressed further in our discussion.

**Human experiments: Movie Watching:**

For the movie watching data, we had 336 ± 16 working channels. We retained 615±12 s of data free from artifacts. Figures 6A and 6B show the spatial distribution of relative alpha band and beta band power across the brain, respectively. Both results have been averaged across 5 experimental repeats. Alpha power was concentrated in the occipital and temporal lobes whilst beta power was maximal over the parietal lobes; in particular the sensory and motor areas. These results are in agreement with well-established findings. Figure 6C shows the power spectral density for six example regions in the brain; in all cases, the black line shows the mean PSD, whilst the grey shaded areas show the standard deviation across 5 repeats of the experiment. Occipital regions show a clear alpha peak in the spectrum, whilst sensorimotor areas have a large beta band component. Again, these results are in good agreement with established literature.

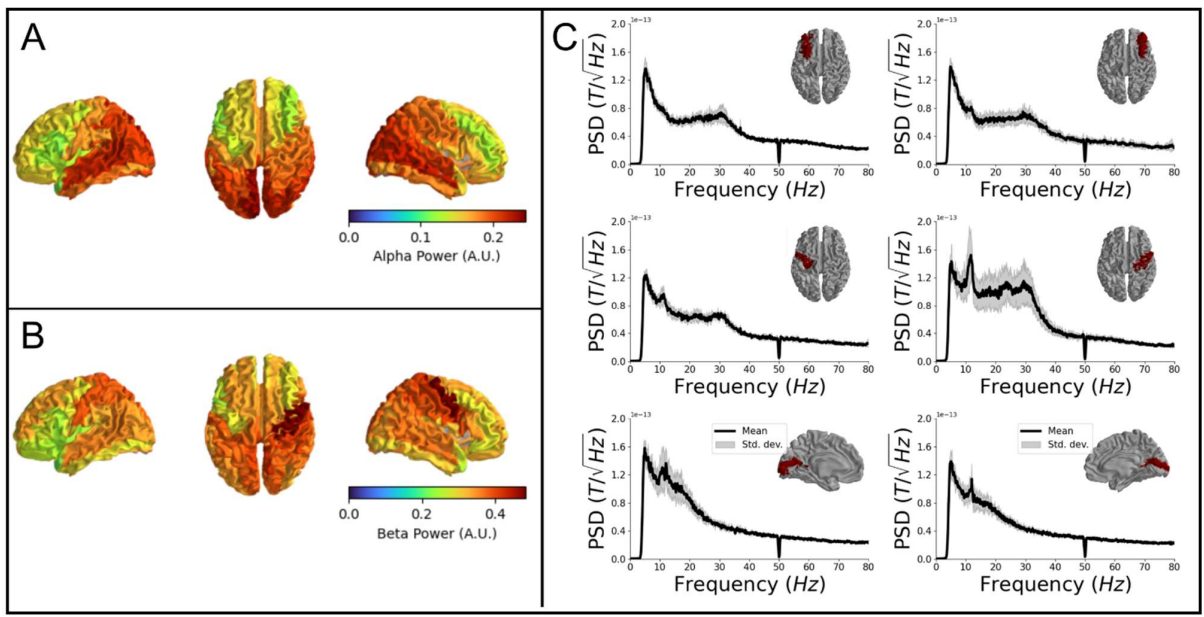

*Figure 6: resting state oscillatory power:* A) Spatial distribution of relative alpha power plotted as a heat map for all AAL regions. B) Spatial distribution of relative beta power. C) Examples of power spectral density reconstructed for 6 AAL regions – left and right middle frontal gyrus, primary motor cortex and primary visual cortex. Note the differences in power spectrum in different brain regions.

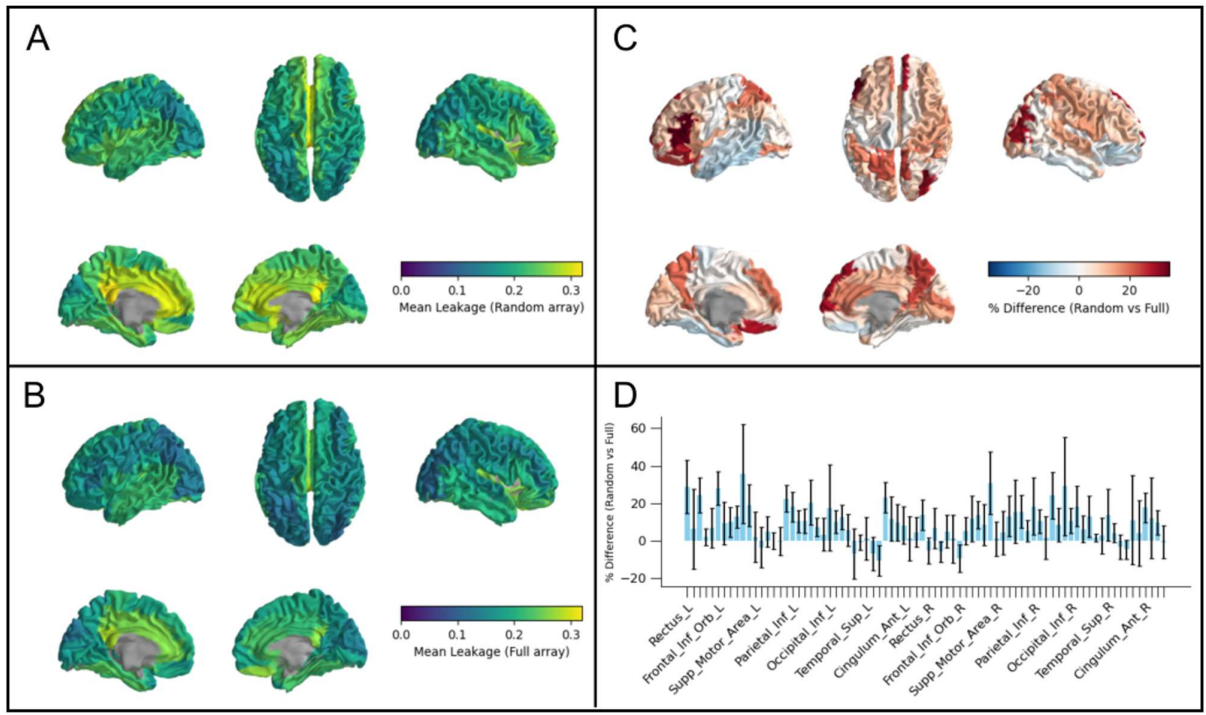

*Figure 7: The advantages of high channel count for spatial resolution:* A) Signal leakage from all AAL regions for a randomly sampled array with 192 channels. B) Signal leakage for a full 336 ± 16-channel array. C) The difference in signal leakage between the random and full arrays, plotted as a percentage (i.e. 100((random – dense)/dense)). Note that for most brain regions the difference is positive meaning that the dense array is advantageous in terms of signal leakage. D) Bar chart showing relative improvement (again as a percentage) in spatial resolution for all brain regions. Here, the bar height represents the mean for each region and the error bar shows standard deviation across 5 experimental runs.

Figure 7 shows the advantages of high channel-count for spatial resolution. In panels A and B, the colours represent the magnitude of signal leakage from a single region to all other regions in the brain. Larger values indicate more leakage, and therefore a lower spatial resolution. Leakage tends to be smaller for regions with good sensor coverage and worse for deeper regions as would be expected. However, the leakage is generally higher (lower spatial resolution) for the sparse array (Figure 7A) than the dense array (Figure 7B). This is also shown in Figure 7C which delineates the percentage difference between leakage measured for the sparse and dense array. Note that the values in Figure 7C are positive for most brain regions, indicating that the addition of extra channels improves spatial resolution. This is further shown in the bar chart in Figure 7D which shows the relative improvement (again measured as a percentage) afforded by moving from the sparse to the dense array for each region. Note that in some brain regions the addition of extra channels improves the spatial resolution by >30%.

**DISCUSSION**

OPM-MEG is proving to be a transformative shift in MEG instrumentation. However, its ultimate adoption to replace SQUID-based devices will depend on our ability to build high channel count systems (beyond the ~300 used for conventional MEG) that will maximise sensitivity, spatial resolution and coverage.

Such high-density arrays must offer characterisation of magnetic field distributions with a high degree of accuracy, whilst also maintaining the advantages – such as motion robustness and lifespan compliance – that have become important reasons for the success of OPM-MEG to date. In this paper, we have constructed a high-channel-count OPM-MEG device using triaxial sensors and a synchronised integrated miniaturised electronic control system. The platform itself is flexible, with the electronics able to expand (or contract) to support any arbitrary number of sensors. We have demonstrated a new fast method for calibration of the system which ensures accuracy of measurements, and via both phantom and human experiments, we have demonstrated viability as a platform for electrophysiological imaging.

The initial validation of our system was undertaken using a PCB-based phantom. This enables a means to gather data where the ground truths – in this case the relative locations of the 5 magnetic dipoles – are known with a high degree of accuracy (from the PCB manufacturing process). Localising dipoles in this way represents an important test of any MEG system, since accurate localisation can only be achieved if the array offers high fidelity recording of a magnetic field distribution. We showed that, following calibration, we could determine the locations of the dipoles to an accuracy better than 1 mm. Furthermore, we showed that the correlation between the magnetic fields measured by the array, and those expected from the dipoles, agreed with a correlation coefficient >0.998.

This is an important result for four reasons: First, it *validates the sensors and the process used to calibrate them*. It is noteworthy that the calibration (via the matrix coil) and the validation method (using the phantom) are independent; whilst both work by imposing well-characterised fields on the array, those fields are very different. The fact that we obtain such good agreement between the measurement and the phantom can only mean that the calibration is working. Indeed, data shown in Figure 3 shows that without calibration, the agreement between the measured and modelled fields is degraded. Second, the high level of agreement suggests that our system is *not subject to systematic errors due to crosstalk*. Crosstalk affects both the gain and orientation sensitivity of a sensor, which in turn can affect localisation accuracy. It can be corrected by alternative designs of on-board coil (Nardelli et al., 2019) but these designs typically add bulk to sensors and are therefore limiting for practical high-density arrays. For open loop sensor operation, as long as the relative locations and orientations of the sensors remain constant during a recording, then the effect of cross talk will also remain constant and is accounted for by calibration. The high degree of accuracy demonstrated by our phantom experiment would be unlikely if our array was subject to crosstalk errors. Third, any future use of OPM-MEG systems will *require quality assurance (QA) processes that are accurate and practical*. This is important for all MEG studies but becomes particularly important when using OPM-MEG clinically (for example, to determine the location of epileptic foci (Rampp et al., 2019)). Here our phantom offers a practical platform to enable this. Finally, *our calibration process is both simple and practical* compared to previous studies (Hill et al., 2025). A limitation of calibration using an external coil is that, if the array moves during the process, then the accuracy of the result can be degraded. However, here our calibration process requires just 1-s of usable data (see Appendix). This should be possible to acquire even in the case of difficult participants (e.g., children) who find it hard to keep still, since data could be recorded for a longer duration alongside head tracking (which is routine during many OPM-MEG studies). The calibration itself could then be extracted from the 1-s window with the smallest subject motion.

Our visual experiment demonstrated the utility of a high-density device to detect visual gamma oscillations. Results showed that stimulus induced modulation of gamma activity localised to the primary visual areas. Further, TFS data showed a narrow band gamma response which was present during presentation of the static grating and became slightly stronger during presentation of the drifting grating. These findings are in good agreement with previous literature (Fries et al., 2008; Hill et al., 2024; Muthukumaraswamy, 2013). Moreover, the gamma experiment showed the advantages of high-density arrays. Our previous work used theory, simulation, and experiments to demonstrate a linear relationship between the total signal measurable by an array ($\|l\|$) and the SNR of the resulting beamformer reconstructed signal. Here, our results followed the same relationship, with gamma SNR again improving linearly with $\|l\|$ which was increased by adding channels up to a maximum of 350. Between a 174-channel (mean) array and 349-channel (mean) array, $\|l\|$ increased by a factor of ~1.4 and SNR increased by a factor of ~1.3. However, we stress that increasing the total channel count is not the only way to increase $\|l\|$. In cases where the brain region of interest is known a-priori it is possible to achieve the same sensitivity benefits by simply redistributing a smaller number of sensors over a smaller area of the scalp. For example, data in Figure 5 suggest that via judicious sensor placement, one can reach 95% of the SNR of a 384-channel array using just 177-channels. This is one of the significant advantages of OPMs over SQUIDs and shows that high channel count isn't always required (for example, in cases where a particular task is known to activate a specific region, or potentially in some clinical scenarios such as epilepsy, where seizure semiology may provide an a-priori region for further investigation). However, in cases where whole brain coverage is required with high sensitivity, then obviously high channel counts become a necessity.

The movie watching data also demonstrate the utility of our high-density system, with results in Figure 6 showing the expected distribution of alpha and beta power across the occipital and parietal lobes respectively (e.g. see (Hunt et al., 2016) for a comparison with conventional MEG). Moreover, the movie watching results show the utility of a higher channel count to improve spatial resolution. We chose to characterise spatial resolution in terms of the leakage between regions across the AAL parcellation. Such leakage is a function of 1) the correlation between forward field patterns from two regions and 2) the SNR of the reconstruction for each region. Forward field correlation is not impacted by sensor count (though it is reduced significantly by moving sensors closer to the scalp – giving OPMs an inherent advantage over SQUIDs (Boto et al., 2019)) However, SNR (as evidenced by our gamma band experiments) is impacted by high sensor density and is also enhanced by accurate calibration (Hill et al., 2025). It is for this reason that we likely see decreased signal leakage with higher numbers of channels. This will be important in future studies of whole brain network connectivity.

The present study has several limitations which should be noted. The form factor of both the helmet and electronics requires work. The helmet housing the sensors was a simple bespoke 3D printed structure. It ensured that sensors were positioned in close proximity to the scalp. However, the sensors themselves are heated, the helmet becomes warm during operation, and the present design did not provide any active cooling to the sensors. To prevent discomfort to the participant, the sensor array was run for a single experiment and rebooted between experiments and allowed to cool down. This was sufficient for a proof of principle demonstration. However, if high density arrays are to be used for long periods (e.g., when trying to detect epileptic spikes in patients, then it is likely that helmets will need to incorporate either improved insulation or active cooling (e.g., by passing air (Pang et al., 2022) or water across the sensors to efficiently remove the heat that they generate). An additional concern of high-density arrays is that the combined weight of the sensors becomes large and makes wearable arrays impractical – particularly for use in children who may find it difficult to support the weight on their heads. However, here the total weight of the 128 sensors was 512 g and the total helmet weight (when populated with sensors) was 974 g. This was comfortably supported by the participant. Similarly, the form factor of the electronics in the present study was not optimised and simply comprised two separate electronics boards which were independently powered and synchronised by a separate signal generator. The electronics boards were each 0.36 x 0.20 x 0.06 cm and weighed 0.81 kg each. The two boards could therefore easily be integrated into a single package which could be mounted within the patient support or even worn by the participant as a backpack to enable ambulatory studies. Finally, in the present study we had a relatively large number of channels that were eliminated from our final analyses. This was caused by two separate issues; firstly, a few sensors were among the first triaxial OPMs built. They were handmade (rather than manufactured) and consequently they are less robust than newer sensors. Secondly, some sensors used a ribbon cable which attaches via a small clamp on the top of the sensor; as sensors age, this can become loose and consequently sensors can became detached during a scan. Both of these issues are solved with more recent sensor designs (the latter problem being solved by a cylindrical (rather than ribbon) cable and a push-fit connector (rather than a clamp)).

**CONCLUSION**

We have successfully constructed an OPM-MEG array with >300 channels, using triaxial sensors and integrated miniaturised electronic control units. We have developed and applied new calibration methods to ensure the that the fields measured are high fidelity and we have introduced a phantom that provides quantification of that high fidelity. We have demonstrated our system by measuring MEG data during a visual task and a movie watching paradigm, and we have demonstrated the advantages of high channel count for sensitivity and spatial resolution. The introduction of the first OPM-MEG system with a channel count higher than that of SQUID-based devices is a significant step on the path towards OPMs becoming the sensor of choice for MEG measurement.

**APPENDIX: FURTHER COIL CALIBRATION RESULTS**

The paper describes a calibration technique in which matrix coils are used generate uniform fields and linear gradients, and a fitting algorithm is used to determine the sensor positions, orientations and gains. To further investigate performance of this method, we undertook two additional experiments. First, we aimed to compare our "gradient fitting" calibration to a more established "coil-by-coil" method (originally described in (Hill et al., 2025)). Second, we aimed to determine how the amount of data used for the gradient fitting approach affected the overall accuracy of the calibration.

**Method:**

**Calibration comparison:**

The coil-by-coil method has been described in detail in a previous paper. Briefly, a forward model describing the field per unit current generated by each matrix coil element was created, within a volume of interest across the centre of the MSR. (This was done once, using a flux gate magnetometer to accurately characterise the fields). To perform a calibration, we generated sinusoidally modulated fields, simultaneously using 96 coil elements, with each element generating a different frequency (between 2 and 11.5 Hz with a separation of 0.1 Hz). We captured 20 s of data from all OPM channels in the array. A Fourier approach was used to isolate signals from each coil element, giving the contribution of each coil to data recorded at each OPM-MEG channel. By fitting these data to the forward model, it becomes possible to determine the location, orientation and gain of the channel. This can then be repeated for all channels.

To compare the coil-by-coil and gradient approaches, we mounted 64 triaxial sensors in a 3D printed OPM-MEG helmet (Cerca Magnetics Ltd., Nottingham, UK). The helmet was placed close to the centre of the MSR and we ran both calibration procedures 6 times. We extracted sensor positions using each technique and computed a matrix of the distances between each possible sensor pair in the array. We then subtracted these matrices from the expected values from the CAD model of the helmet to produce residual matrices showing the 'error' in calibration derived distances (i.e. taking the CAD as the ground truth). In addition, we took the calibration derived gain values from the coil-by-coil method and plotted them against equivalent values from the gradient approach.

**Quantity of data:**

For the coil-by-coil method to work, we required a relatively long data segment (20 s in this case), to get the spectral resolution required in Fourier space to separate out signals from 96 coil elements. However, using the gradient method, we require just 8 signals (three uniform fields and 5 gradients). Consequently, the data segment required can be much shorter. To investigate how the amount of data (i.e. the speed at which the gradient method can be run) affects the accuracy of the calibration, we took the data from the experiment described above (which used a 4-s data segment) sequentially reduced the amount of data (to 1 s in steps of 0.5 s) used for the calibration process.

**RESULTS:**

Figure A1A presents the error matrices; the difference between the Gradient method and CAD is shown in the upper panel; the difference between the coil-by-coil method and CAD is in the lower panel. In both cases data are averaged across all 6 runs of the experiment. The average residuals across all sensor pairs were 1.3 ± 0.9 mm and 2.0 ± 0.8 mm (mean ± standard deviation across all matrix elements) for the gradient and coil-by-coil calibrations respectively. The average residual between the two calibrations was 2.1 ± 0.7 mm. These results indicate good agreement between the two calibration methods and the ground truth (CAD) and good agreement with each other. Figure A1B shows the channel gains derived from two calibrations plotted against each other (data for all 6 experimental runs are shown on the same plot). There is a clear linear relationship indicating good agreement between approaches.

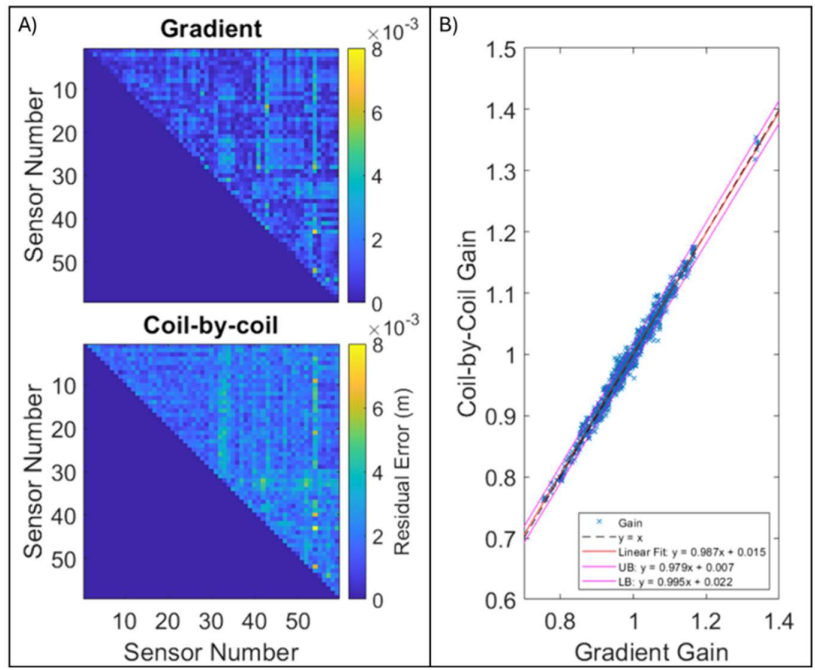

*Figure A1: Comparison of calibration techniques:* A) Matrices showing the mean (across experimental repeats) sensor-to-sensor distance error between CAD and calibration. The gradient method is shown in the upper panel and coil-by-coil method is shown in the lower panel. B) Channel gains determined by the gradient and coil-by-coil calibrations, plotted against each other (all channels and all runs overlaid). The black dashed line shows the line of equality; the red line shows a linear fit. The close relationship suggests that the two calibration methods result in very similar gain measures.

In Figure A2, the left-hand panel shows the average distance error in gradient calibration derived sensor positions (compared to CAD), as a function of the total amount of data used for the calibration fit. The right-hand panel shows the difference in gain between the gradient and coil-by-coil calibrations, again as a function of the amount of data used for the gradient calibration method. In both cases, the line plot shows the mean and the error bar shows standard deviation across 6 experimental repeats. These results show that, as may be expected, errors increase with decreasing data length (an effect that may be due to the spacing of peaks being just 1 Hz which could be optimised further in the future). However, this effect is marginal and, even for just 1 s of data, the residual mean is just 2.5 mm with gain errors <1.5%.

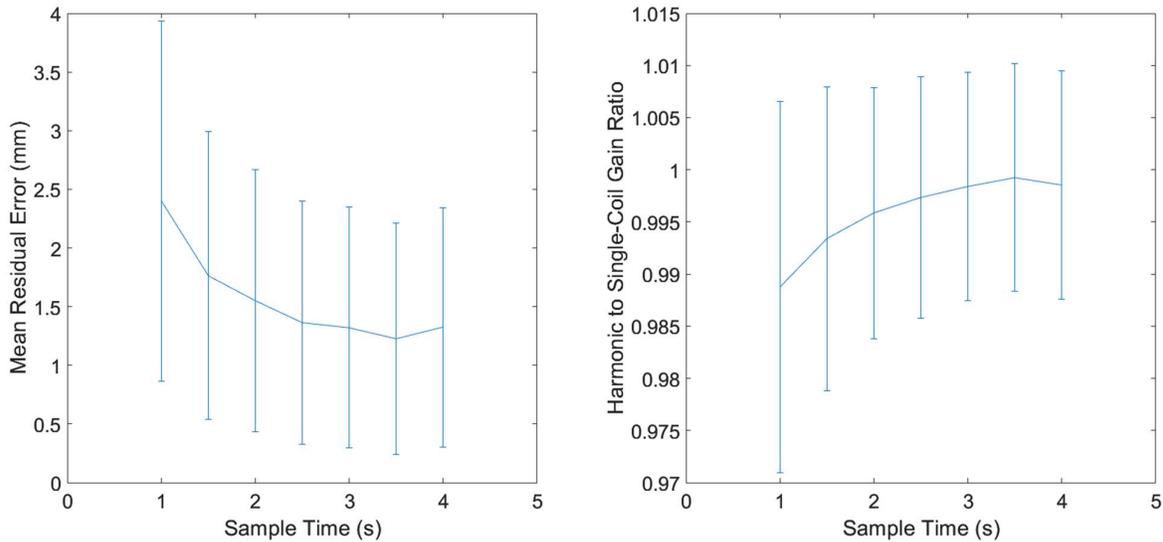

*Figure A2: Accuracy as a function of sample time: Left: the average error between gradient calibration and CAD, plotted as a function of the amount of data used for the calibration (sample time). Right: the average ratio of the gradient and coil-by-coil derived gains, plotted as a function of sample time. The error bars show the standard deviation across runs.*

**DISCUSSION**

This additional experiment shows good agreement between our new "gradient" calibration method and a more established technique. Moreover, it shows that accurate calibration can be achieved using just 1 s of data, meaning that gradient-based calibration can more readily be employed in realistic situations. Specifically, one disadvantage of matrix-coil based calibrations is that if the head moves during the calibration process, the sensors see a changing field and so the accuracy of the resulting sensor locations, orientations and gains is diminished. Expecting participants (especially children or patients) to sit still for a 20-s data collection is unreasonable. However, with accurate results achievable in just 1 s, the technique becomes much more practical. For example, even in participants who find it very hard to keep still, the calibration process could be run at multiple intervals and, if paired with head tracking, experimenters could isolate the 1-s window with the lowest head movement. Multiple rapid calibrations could also be used to assess sensor stability at regular intervals. In sum, we believe this rapid and simple calibration method could be a useful addition to OPM-MEG systems.

**ACKNOWLEDGEMENTS**

This work was supported by an Engineering and Physical Sciences Research Council (EPSRC) Grant (EP/Z535722/1) and the UK Quantum Technology Hub in Sensors Imaging and Timing (QuSIT), funded by EPSRC (EP/Z533166/1). We acknowledge a Medical Research Council (MRC) Mid-Range Equipment grant (MC_PC_MR/X012263/1).